\documentclass[%
pra,
twocolumn,
groupedaddress,
showpacs,preprintnumbers,
amsmath, amssymb, aps,raggedbottom
]{revtex4-1}
\usepackage{color}
\usepackage{amsmath}
\usepackage{amssymb}
\usepackage{graphicx}
\usepackage{epsfig}
\usepackage{epstopdf}
\usepackage{dcolumn}
\usepackage{bm}
\usepackage{natbib}
\usepackage{textcomp}

\begin{document}

\title{Supersymmetry-generated complex optical potentials with real spectra}

\author{Mohammad-Ali Miri}\email{miri@knights.ucf.edu} \author{Matthias Heinrich} \author{Demetrios N. Christodoulides}
\affiliation{CREOL$/$College of Optics, University of Central Florida, Orlando, Florida, USA}
\begin{abstract}
We show that the formalism of supersymmetry (SUSY), when applied to parity-time ($\mathcal{PT}$) symmetric optical potentials, can give rise to novel refractive index landscapes with altogether non-trivial properties. In particular, we find that the presence of gain and loss allows for arbitrarily removing bound states from the spectrum of a structure. This is in stark contrast to the Hermitian case, where the SUSY formalism can only address the fundamental mode of a potential. Subsequently we investigate isospectral families of complex potentials that exhibit entirely real spectra, despite the fact that their shapes violate $\mathcal{PT}$-symmetry. Finally, the role of SUSY transformations in the regime of spontaneously broken $\mathcal{PT}$ symmetry is investigated.
\end{abstract}


\maketitle
\section {INTRODUCTION}

Supersymmetry (SUSY) was originally conceived within the framework of quantum field theories and high-energy physics \cite{Ram1971,Nev1971,Gel1971,Vol1973,Wes1974}. Since then, aspects of SUSY have been systematically employed in many and diverse areas of physics and mathematics, including nonrelativistic quantum mechanics \cite{Wit1981,Coo1983,Ste1985_1,*Ste1985_2,Kha1989,Gen1985,Lah1990,Coo1995}. In particular, SUSY techniques have been instrumental in identifying analytically solvable potentials, the investigation of shape invariance as well as the development of powerful approximation methods \cite{Coo1995}. In the context of nonrelativistic quantum mechanics, SUSY is established by factorizing the Schr\"odinger equation in order to construct superpartner Hamiltonians. The potentials corresponding to this pair of Hamiltonians then share the same eigenvalue spectrum, with the exception of the ground state. If however this ground state is present in the spectra of both partner potentials, as indicated by a vanishing Witten index \cite{Wit1981}, SUSY is then said to be broken. Along similar lines, parametric families of Hamiltonians sharing the exact same eigenvalue spectrum - including the ground state - can be constructed from a given potential \cite{Kha1989,Coo1995}. Quite recently, we have shown that some of the fascinating applications of SUSY can be explored and utilized within the field of optics \cite{Miri2013}. In particular, it was demonstrated that supersymmetry can establish perfect phase matching conditions between a great number of modes, thus enabling selective mode filtering applications. In addition, it was shown that optical structures related via SUSY can exhibit identical reflectivities and transmittivities irrespective of the angle of incidence - even under strong index contrast conditions.

On the other hand, in the past decade or so, non-Hermitian systems have been a subject of intense research \cite{Ben1998,*Ben2002,*Ben2007,Lev2001,*Ahm2001,*Ahm2005,Mos2002_a,*Mos2002_b}. Interest in such settings was sparked by the pioneering work of Bender and Boettcher, who showed that a wide range of complex Hamiltonians can exhibit entirely real spectra, provided they are invariant under a simultaneous reversal of parity and time - i.e. they obey $\mathcal{PT}$ symmetry \cite{Ben1998}. In general, the complex potentials involved in $\mathcal{PT}$-symmetric Hamiltonians must fulfill the condition $V^*(X)=V(-X)$. In the context of quantum mechanics, efforts were undertaken to extend the standard intertwining relations of SUSY-QM to the complex domain of non-Hermitian Hamiltonians \cite{Can1998,And1999,Bag2000,Zno2000,Dor2001,Can2001,Lev2002,Ber2011}. Interestingly, it was found that such systems might exhibit a zero Witten index, even if SUSY is unbroken \cite{Zno2000}. Furthermore, Darboux transformations have been utilized for constructing complex potentials that display real spectra \cite{Can1998}.

Recently it was noted that non-Hermitian, and in particular $\mathcal{PT}$-symmetric, Hamiltonians can be realized in optics \cite{Elg2007,*Mak2008,*Mus2008}. To this end, optical gain and loss can be judiciously incorporated in the refractive index distribution of a system as a means to construct complex optical potentials \cite{Reg2012,Rut2010,Guo2010,Fen2013}. It soon became apparent that $\mathcal{PT}$-symmetry can enable effects and behavior that would have been otherwise impossible in conventional optical structures. These include band merging, double reflection, breakdown of the left-right symmetry, the abrupt transition from lasing to absorbing modes, and mode selection in laser amplifiers, to mention a few \cite{Elg2007,*Mak2008,*Mus2008,Reg2012,Rut2010,Guo2010,Fen2013,Kla2008,Lon2009,*Lon2010,Sch2010,Lin2011,*Kul2005,Suk2010,*Mir2011,Cho2011,*Lie2012,Miri2012_c,Miri2012_b,Jog2011,*Vem2011,Mid2010,Gra2011,Sza2011,Elg2012,Zez2012,Miri2012_a}. Clearly of interest will be to extend the domain of such complex optical potentials beyond the constraints of $\mathcal{PT}$ symmetry.

In this work we explore the optical ramifications of supersymmetry in the context of complex refractive index landscapes. We show that the SUSY formalism allows for the construction of partner structures where the fundamental mode, or any other higher order guided mode, can be removed at will. Starting from a $\mathcal{PT}$-symmetric configuration, we then investigate isospectral families of non-Hermitian index landscapes that share the exact same eigenvalue spectrum. Through this approach, one can synthesize optical structures where the guided modes experience zero net gain and loss despite of the fact that their shape violates $\mathcal{PT}$ symmetry. Finally, refractive index profiles with spontaneously broken $\mathcal{PT}$ symmetry are investigated. In this case it is shown that removing the resulting pair of complex conjugate modes by means of SUSY leads to a $\mathcal{PT}$-symmetric waveguide without a spontaneous symmetry breaking.

\section {SUSY IN $\mathcal{PT}$-SYMMETRIC OPTICAL POTENTIALS}

Let us first consider how the notion of supersymmetry can be applied in complex optical potentials. As previously shown \cite{Miri2013}, the SUSY formalism can be generally used in arbitrary one-dimensional refractive index landscapes. In fact, this is the case even under high-contrast conditions where the degeneracy between TE and TM waves is broken and necessitates the use of the Helmholtz equation \cite{Miri2013,Koc2005}. Here, for brevity, we limit our scope to one-dimensional weakly guiding settings. In this regime, the beam dynamics can be described within the paraxial approximation. In our system, $n(x)=n_0+\Delta n(x)$ describes the refractive index distribution in the transverse coordinate $x$, where the index modulation $\Delta n(x)$ is assumed to be weak compared to the background index $n_0$, $|\Delta n(x)| \ll n_0$. Under these conditions one finds that the slowly varying envelope $U$ of the electric field component $E(x,z)=U(x,z)e^{ik_0 n_0 z}$ satisfies the following evolution equation:
\begin{equation}
\label{eq1}
i\frac{\partial U}{\partial Z}+\frac{\partial^2U}{\partial X^2}+V(X)U=0.
\end{equation}
Here the normalized transverse and longitudinal coordinates are respectively given by $X=x/x_0$ and $Z=z/(2k_0 n_0 x_0^2 )$, where $x_0$ is an arbitrary length scale, and $k_0=2\pi/\lambda_0$ is the wave number corresponding to the free space wavelength $\lambda_0$. The optical potential $V(x)$ is directly proportional to the refractive index variation,
\begin{equation}
\label{eq2}
V=2k_0^2 n_0 x_0^2 \Delta n(x),
\end{equation}
and in general is complex, $V=V_R+iV_I$, where the real part $V_R(X)$ is the outcome of an index modulation, while the imaginary part $V_I(X)$ indicates the presence of gain or loss.

Looking for stationary (modal) solutions of the form $U(X,Z)=\psi(X)e^{i\mu Z}$, we then obtain the following Schr\"odinger eigenvalue problem:
\begin{equation}
\label{eq3}
H\psi=-\mu \psi,
\end{equation}
where the operator $H=-d^2/dX^2-V(X)$ represents the Hamiltonian of the optical configuration and $\mu$ the respective eigenvalue. 
We now assume that a given potential $V^{(1)}$ supports at least one guided optical mode $\psi_1^{(1)}(X)$ with a corresponding eigenvalue $\mu_1^{(1)}$. Following the approach detailed in \cite{Coo1995}, one can then factorize the Hamiltonian as $H^{(1)}+\mu_1^{(1)}=BA$ with
\begin{subequations}
\label{eq4} 
\begin{eqnarray}
A=+\frac{d}{dX}+W, \label{eq4a}
\\
B=-\frac{d}{dX}+W. \label{eq4b}
\end{eqnarray}
\end{subequations}
Note that, whereas in Hermitian systems described by a real-valued superpotential $W(X)$ the two operators $A$,$B$ form a Hermitian-conjugate pair, this is no longer true in the general case of a complex $W$ ($B\neq A^\dagger$).

Defining a partner Hamiltonian as $H^{(2)}+\mu_1^{(1)}=AB$, one quickly finds that the optical potentials of the original and the partner system can both be generated from the superpotential and its first derivative:
\begin{equation}
\label{eq5}
V^{(1,2)}(X)=\mu_1^{(1)}-W^2\pm W'.
\end{equation}
It readily follows that the two optical potentials $V^{(1,2)}$ then share a common set of eigenvalues \cite{Coo1995}:
\begin{equation}
\label{eq6}
\mu_m^{(1)}=\mu_{m-1}^{(2)} ~~ , ~~ m>1.
\end{equation}

\begin{figure}[b]
\begin{center}
\includegraphics[width=1\linewidth]{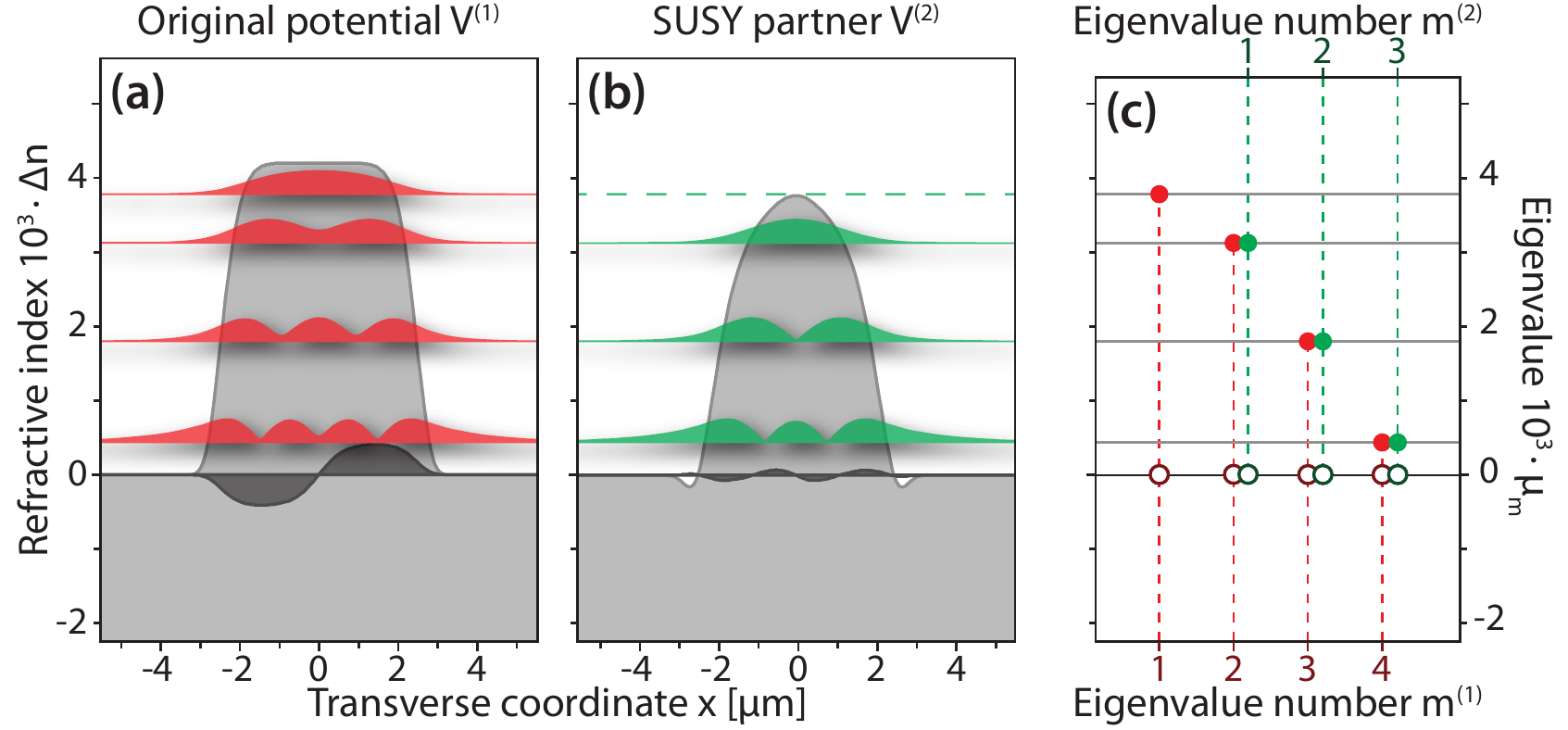}
\caption{(Color online). (a) Refractive index profile (real part: light gray, imaginary part: dark gray area) of a $\mathcal{PT}$-symmetric multimode waveguide supporting a total of four bound states (shown absolute values $\left|\psi_m^{(1)}\right|$ at the vertical positions corresponding to their respective eigenvalues $\operatorname{Re}\left(\mu_m^{(1)}\right)$. (b) Corresponding SUSY partner and its three modes. (c) Eigenvalue spectra of the two structures $\operatorname{Re}\left(\mu_m^{(1,2)}\right)$ are shown as full circles, whereas empty circles denote $\operatorname{Im}\left(\mu_m^{(1,2)}\right)$.}
\label{fig1}
\end{center}
\end{figure}

\noindent The only exception is the fundamental mode of $V^{(1)}$, which lacks a counterpart in $V^{(2)}$. Note that this SUSY mode partnership is not limited to the discrete sets of bound states, but rather extends to the continua of radiation modes of both structures. The operators $A$ and $B$ also provide a link between the wave functions of the two potentials:
\begin{subequations}
\label{eq7} 
\begin{eqnarray}
\psi_m^{(2)}=A\psi_{m+1}^{(1)}, \label{eq7a}
\\
\psi_{m+1}^{(1)}=B\psi_m^{(2)}. \label{eq7b}
\end{eqnarray}
\end{subequations}
In order to derive an expression for the superpotential, we make use of the fact that $A$ should annihilate the fundamental mode of the first potential; $A\psi_1^{(1)}=0$. Therefore, by using Eq.~\eqref{eq4a}, $W$ can be written as a logarithmic derivative of the fundamental mode's wave function:
\begin{equation}
\label{eq8}
W=-\frac{d}{dX}\ln{\left (\psi_1^{(1)}\right)}.
\end{equation}
Similarly, the partner potential $V^{(2)}$ can be expressed in terms of $V^{(1)}$ and $\psi_1^{(1)}$ as follows:
\begin{equation}
\label{eq9}
V^{(2)}=V^{(1)}+2\frac{d^2}{dX^2}\ln{\left (\psi_1^{(1)}\right)}.
\end{equation}
We now apply this formalism when $V^{(1)}$ is $\mathcal{PT}$-symmetric, i.e. $V^{(1)}(-X)={\left (V^{(1)}(X)\right )}^*$. At this point we also assume that the $\mathcal{PT}$ symmetry of $V^{(1)}$ is not broken. Under these conditions, the eigenvalue spectrum is real-valued, i.e. $\operatorname{Im}{\left (\mu_m^{(1)}\right )}=0$, and the individual modes inherit the potential's symmetry: $\psi_m^{(1)}(-X)={\left (\psi_m^{(1)}(X)\right )}^*$. Following Eq.~\eqref{eq8}, one then concludes that the superpotential should be anti-$\mathcal{PT}$-symmetric: $W^*(X)=-W(-X)$. On the other hand, Eq.~\eqref{eq9} clearly shows that $V^{(2)}$ again respects the condition of $\mathcal{PT}$ symmetry. Since SUSY dictates that its spectrum is also real-valued, it follows that $\mathcal{PT}$ symmetry is unbroken in the partner potential.

Figure 1 illustrates the implications of supersymmetry when for example a $\mathcal{PT}$-symmetric multimode waveguide is considered, that has the refractive index profile
\small
\begin{equation}
\label{eq10}
\Delta n^{(1)}(x)=\delta \left(1+i\gamma\tanh{\left(\frac{x}{w_I\lambda_0}\right)}\right)\exp{\left(-{\left(\frac{x}{w_R\lambda_0}\right)}^8\right)}.
\end{equation}
\normalsize
Here, the index elevation is $\delta=4.2\times 10^{-3}$, the imaginary (gain-loss) contrast is $\gamma=0.1$, and $w_R=2.5$, $w_I=0.6$ are geometry parameters. This waveguide supports a total of four guided modes at a wavelength of $\lambda_0=1\mu m$. The figure shows the real and imaginary parts of the refractive index profile as well as the absolute value $\left|\psi_m^{(1)}\right|$ of the modal distributions (Fig.~\ref{fig1}(a)). The corresponding superpartner waveguide and its three guided modes are depicted in Fig.~\ref{fig1}(b), and the eigenvalue spectra of both structures are compared in Fig.~\ref{fig1}(c). Note that none of the $\mathcal{PT}$-symmetric modes exhibit any nodes in their intensity profile.

\section{REMOVAL OF HIGHER ORDER MODES}

\begin{figure}[b,t,h]
\begin{center}
\includegraphics[width=1\linewidth]{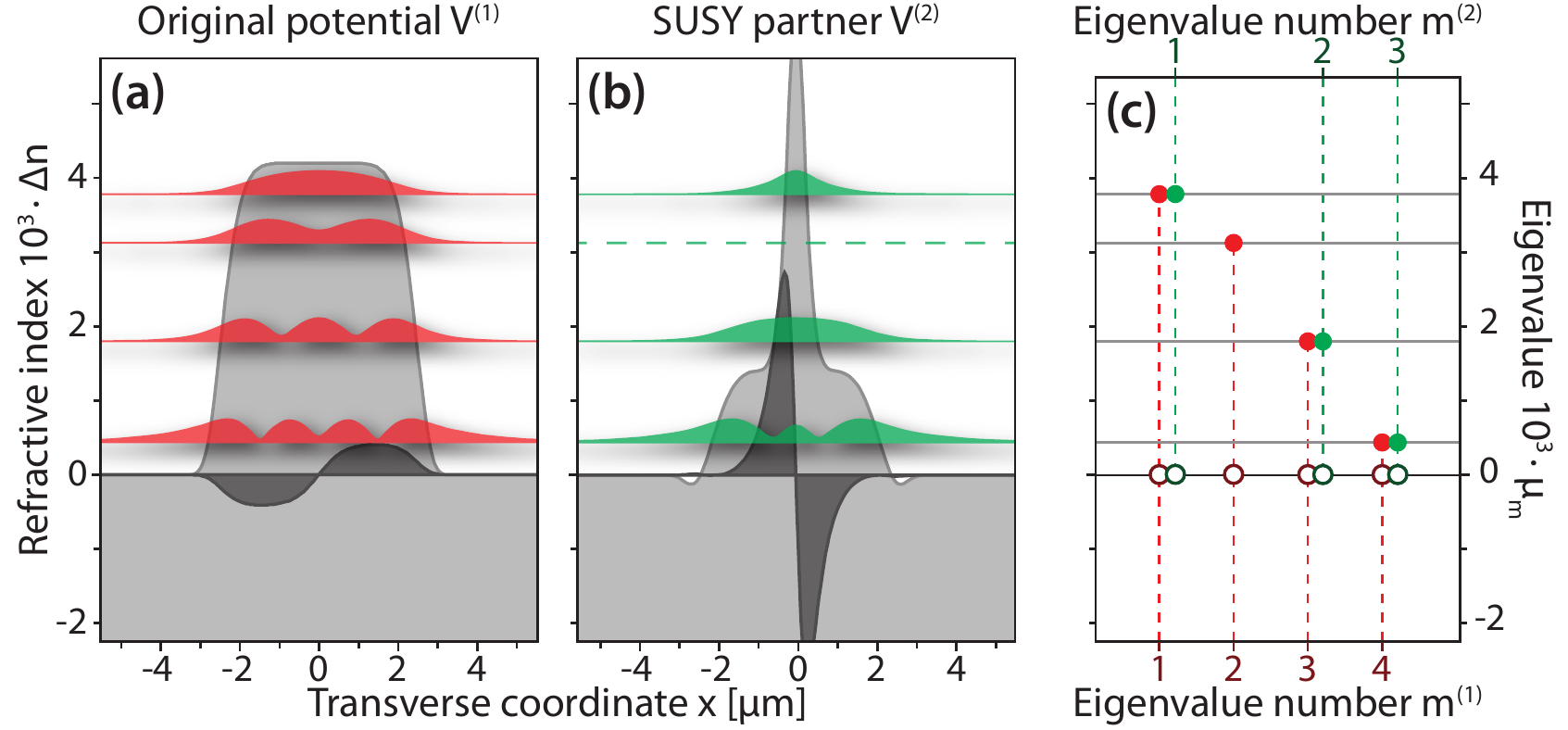}
\caption{(Color online). (a) Refractive index profile of a $\mathcal{PT}$-symmetric multimode waveguide supporting a total of four bound states, as in Fig.~\ref{fig1}. (b) Corresponding SUSY partner where the second mode has been removed from the original waveguide. (c) Eigenvalue spectra of the two structures.}
\label{fig2}
\end{center}
\end{figure}

In Hermitian systems, all modes except for the fundamental state exhibit nodes where the absolute value of the wave function vanishes. Given that the superpotential $W$ as constructed from Eq.~\eqref{eq8} relies on the logarithmic derivative of an eigenfunctions $\psi_m^{(1)}$, in this case one can only use the nodeless ground state $\psi_1^{(1)}$. In contrast, the zeros of the real and imaginary parts of modes associated with non-Hermitian systems do not occur at the same positions. This peculiar behavior now allows one to use any higher order mode $\psi_{m_0}^{(1)}$ ($m_0>1$) (see Fig.~\ref{fig2}(a)) in constructing a SUSY partner, i.e. by removing the eigenvalue $\mu_{m_0}^{(1)}$ from the spectrum. In other words,
\begin{subequations}
\label{eq11} 
\begin{eqnarray}
V^{(1,2)}(X)=\mu_{m_0}^{(1)}-W^2\pm W', \label{eq11a}
\\
W=-\frac{d}{dX}\ln\left(\psi_{m_0}^{(1)}\right), \label{eq11b}
\\
V^{(2)}=V^{(1)}+2\frac{d^2}{dX^2}\ln{\left (\psi_{m_0}^{(1)}\right)}. \label{eq11c}
\end{eqnarray}
\end{subequations}
The relations between eigenvalues and wave functions for these two structures then can be written as
\footnotesize
\begin{subequations}
\label{eq12} 
\begin{eqnarray}
\mu_m^{(1)}=\mu_m^{(2)}; ~ \psi_m^{(2)}=A\psi_m^{(1)}; ~ \psi_m^{(1)}=B\psi_m^{(2)}, ~ m<m_0 ~,\label{eq12a}
\\
\mu_m^{(1)}=\mu_{m-1}^{(2)}; \psi_m^{(2)}=A\psi_{m+1}^{(1)}; \psi_{m+1}^{(1)}=B\psi_m^{(2)}, m>m_0.\label{eq12b}
\end{eqnarray}
\end{subequations}
\normalsize
Figure 2 illustrates the removal of the eigenvalue associated with the second mode from the spectrum of the multimode waveguide discussed in Fig.~\ref{fig2}(a). Again the SUSY partner potential (Fig.~\ref{fig2}(b)) supports three modes, which are now matched to the eigenvalues of the first, third and fourth mode of the original structure. Note that the partner waveguide has been most strongly altered in regions where the removed state had an intensity minimum. There, the second derivative of the wave function's absolute value is maximal, resulting in a pronounced feature in the SUSY partner. In the Hermitian limit $\operatorname{Im}\left(\Delta n^{(1)}\right)\rightarrow 0$, this feature is transformed into a singularity.

\section{REAL SPECTRA WITHOUT $\mathcal{PT}$ SYMMETRY}

In this section we explore the possibility of synthesizing complex potentials, based on SUSY transformations, that support entirely real spectra despite the fact that they violate the necessary condition for $\mathcal{PT}$-symmetry. In the framework of nonrelativistic SUSY quantum mechanics, it is known that one can establish whole families of isospectral potentials that share the spectrum of a given ``parent" potential. Here we will show that this approach can be adapted so as to construct optical systems that happen to be isospectral to a $\mathcal{PT}$-symmetric structure.

\begin{figure}[t,b,h]
\begin{center}
\includegraphics[width=1\linewidth]{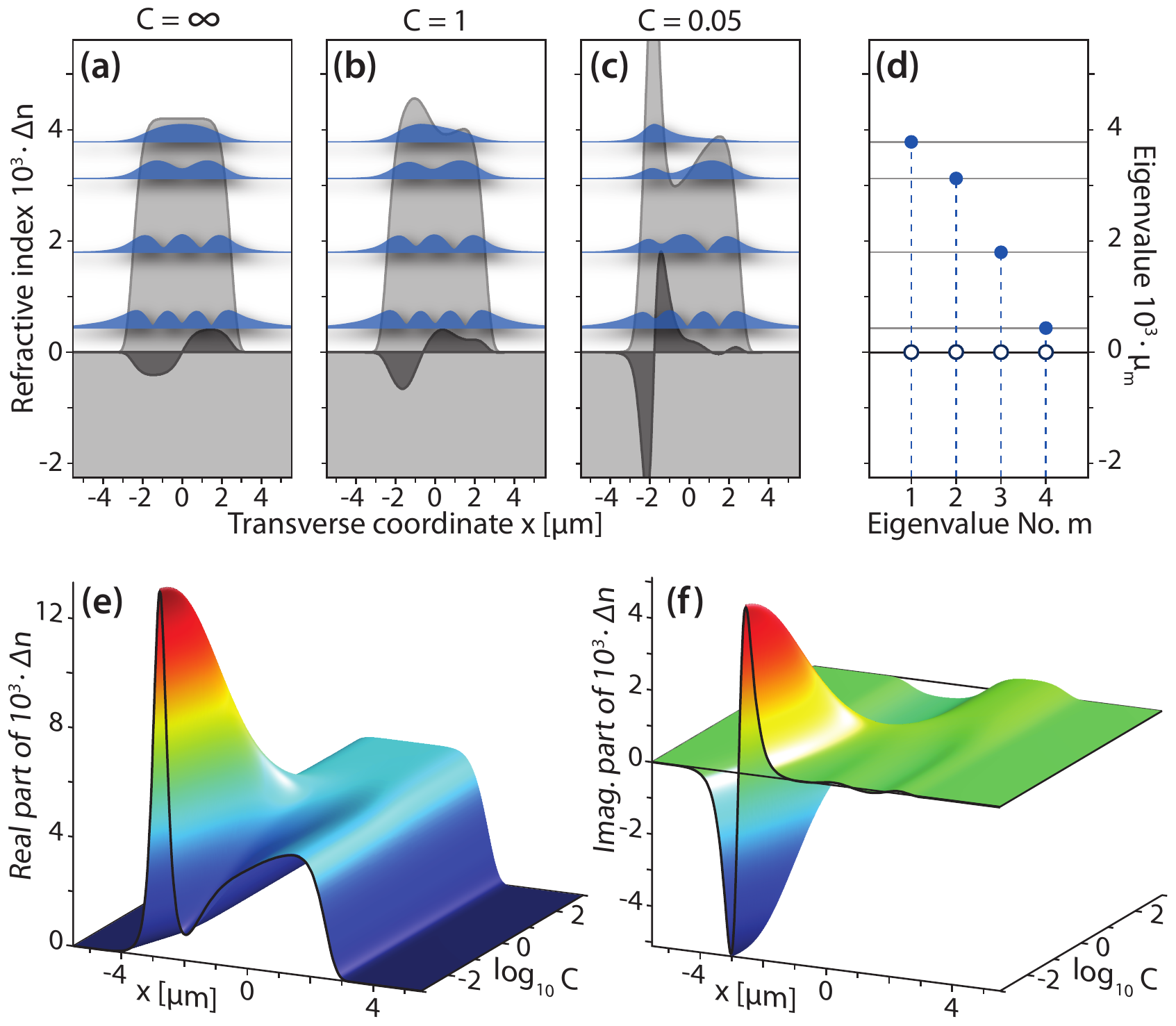}
\caption{(Color online). (a) Refractive index profile of a $\mathcal{PT}$-symmetric multimode waveguide supporting a total of four bound states. For $C\rightarrow\infty$, the parametric family converges toward this parent potential. (b,c) As $C\rightarrow 0$, the potentials and their guided modes become visibly distorted. (d) Regardless, all members of the family share the exact same eigenvalue spectrum. (e,f) Profile of the real and imaginary components of the complex potential associated with this isospectral family after continuously varying $10^{-3}<C<10^{3}$.}
\label{fig3}
\end{center}
\end{figure}

Consider again a $\mathcal{PT}$-symmetric potential supporting at least one guided mode in a complex index profile that satisfies the condition $V^{(1)}(-X)={\left (V^{(1)}(X)\right )}^*$. According to Eq.~\eqref{eq11a}, the superpotential $W$ satisfies the well-known Riccati equation $V^{(2)}(X)=\mu_{m_0}^{(1)}-W^2-W'$. A general solution of this equation $\tilde{W}$ can be written in terms of the particular solution $W$ found in Eq.~\eqref{eq11b} as \cite{Arfken2001} $\tilde{W}=W+1/v$, where $v$ satisfies the first order equation $v'=1+2Wv$. By using $W$, as given in Eq.~\eqref{eq11b}, the solution of this latter equation can be written as $v(X)={\left(\psi_{m_0}^{(1)}(X)\right)}^{-2}\left(C+\int_{-\infty}^{X}{{\left(\psi_{m_0}^{(1)}(X')\right)}^{2}dX'}\right)$, where $C$ is an arbitrary complex constant of integration. This results in the following parametric family of superpotentials:
\begin{equation}
\label{eq13}
\tilde{W}=W+\frac{d}{dX}\ln\left(C+\int_{-\infty}^{X}{{\left(\psi_{m_0}^{(1)}(X')\right)}^{2}dX'}\right),
\end{equation}
and the corresponding isospectral family of complex optical potentials
\small
\begin{equation}
\label{eq14}
\tilde{V}^{(1)}=V^{(1)}+2\frac{d^2}{dX^2}\ln\left(C+\int_{-\infty}^{X}{{\left(\psi_{m_0}^{(1)}(X')\right)}^{2}dX'}\right).
\end{equation}
\normalsize
In order to avoid singular behavior, the parameter $C$ must be appropriately chosen such that the quantity $C+\int_{-\infty}^{X}{{\left(\psi_{m_0}^{(1)}(X')\right)}^{2}dX'}$ is never zero for any $-\infty<X<+\infty$. Note that all the members $\tilde{V}^{(1)}$ of this family form a valid SUSY pair with the same $V^{(2)}$ and are isospectral to $V^{(1)}$. Equation~\eqref{eq14} indicates that in general the members of the isospectral family constructed from the original $\mathcal{PT}$-symmetric potential do not exhibit a $\mathcal{PT}$ symmetric form, i.e. $\tilde{V}^{(1)}(-X)\neq{\left (\tilde{V}^{(1)}(X)\right )}^*$ (see Fig.~\ref{fig3}(a-c)). Nevertheless, as long as $\mathcal{PT}$ symmetry is not spontaneously broken in the parent potential $V^{(1)}$, the spectra of all the members of this family will be entirely real-valued (Fig.~\ref{fig3}(d)).
A closer look at the shape of the respective optical eigenmodes reveals the mechanism behind this unexpected behavior. Even though the gain-loss is no longer antisymmetrically distributed across the waveguide's profile (Fig.~\ref{fig3}(f)), the real part is deformed (Fig.~\ref{fig3}(e)) such that the redistributed mode profiles can maintain a neutral imaginary overlap. To formally justify this intuitive explanation, consider again the eigenvalue equation and its complex conjugate associated with this problem: $\frac{d^2}{dX^2}\tilde{\psi}_m^{(1)}+\tilde{V}^{(1)}\tilde{\psi}_m^{(1)}=\mu_m\tilde{\psi}_m^{(1)}$, $\frac{d^2}{dX^2}{\left(\tilde{\psi}_m^{(1)}\right)}^*+{\left(\tilde{V}^{(1)}\right)}^*{\left(\tilde{\psi}_m^{(1)}\right)}^*=\mu_m^*{\left(\tilde{\psi}_m^{(1)}\right)}^*$. After multiplying these equations by ${\left(\tilde{\psi}_m^{(1)}\right)}^*$ and $\tilde{\psi}_m^{(1)}$ respectively, their difference yields:
\small
\begin{equation}
\label{eq15}
\begin{split}
{\left(\tilde{\psi}_m^{(1)}\right)}^*\frac{d^2}{dX^2}\tilde{\psi}_m^{(1)}-\tilde{\psi}_m^{(1)}\frac{d^2}{dX^2}{\left(\tilde{\psi}_m^{(1)}\right)}^*\\+\left(\tilde{V}^{(1)}-{\left(\tilde{V}^{(1)}\right)}^*\right){\left|\tilde{\psi}_m^{(1)}\right|}^2=\left(\mu_m-\mu_m^*\right){\left|\tilde{\psi}_m^{(1)}\right|}^2.
\end{split}
\end{equation}
\normalsize
Given that the first term represents a total differential, and that the eigenvalue is real $(\mu_m^*=\mu_m)$, we therefore find
\small
\begin{equation}
\label{eq16}
\frac{d}{dX}\left({\left(\tilde{\psi}_m^{(1)}\right)}^*\frac{d}{dX}\tilde{\psi}_m^{(1)}-\tilde{\psi}_m^{(1)}\frac{d}{dX}{\left(\tilde{\psi}_m^{(1)}\right)}^*\right)+2{\left|\tilde{\psi}_m^{(1)}\right|}^2\tilde{V}_I^{(1)}=0.
\end{equation}
\normalsize
Taking also into account that bound states decay exponentially outside the guiding region and vanish at infinity, integration over the entire $X$ axis yields
\begin{equation}
\label{eq17}
\int_{-\infty}^{+\infty}{\tilde{V}_I^{(1)}{\left|\tilde{\psi}_m^{(1)}\right|}^2dX}=0.
\end{equation}
In other words, the overlap integral between the imaginary part of the refractive index profile and the modal intensity always vanishes in such settings. Moreover, a direct integration over the imaginary part of the potential shows that a transformation according to Eq.~\eqref{eq14} does not introduce any changes to the overall gain-loss of the system. Considering that the imaginary part of the $\mathcal{PT}$-symmetric parent potential $V^{(1)}$ itself is anti-symmetric and that $2\operatorname{Im}\left[\frac{d}{dX}\ln\left(C+\int_{-\infty}^{X}{{\left(\tilde{\psi}_{m_0}^{(1)}(X')\right)}^{2}dX'}\right)\right]_{-\infty}^{+\infty}=0$, one also finds
\begin{equation}
\label{eq18}
\int_{-\infty}^{+\infty}{\tilde{V}_I^{(1)}dX}=0.
\end{equation}

\section{SUSY IN STRUCTURES WITH SPONTANEOUSLY BROKEN $\mathcal{PT}$ SYMMETRY}

\begin{figure}[t,b,h]
\begin{center}
\includegraphics[width=1\linewidth]{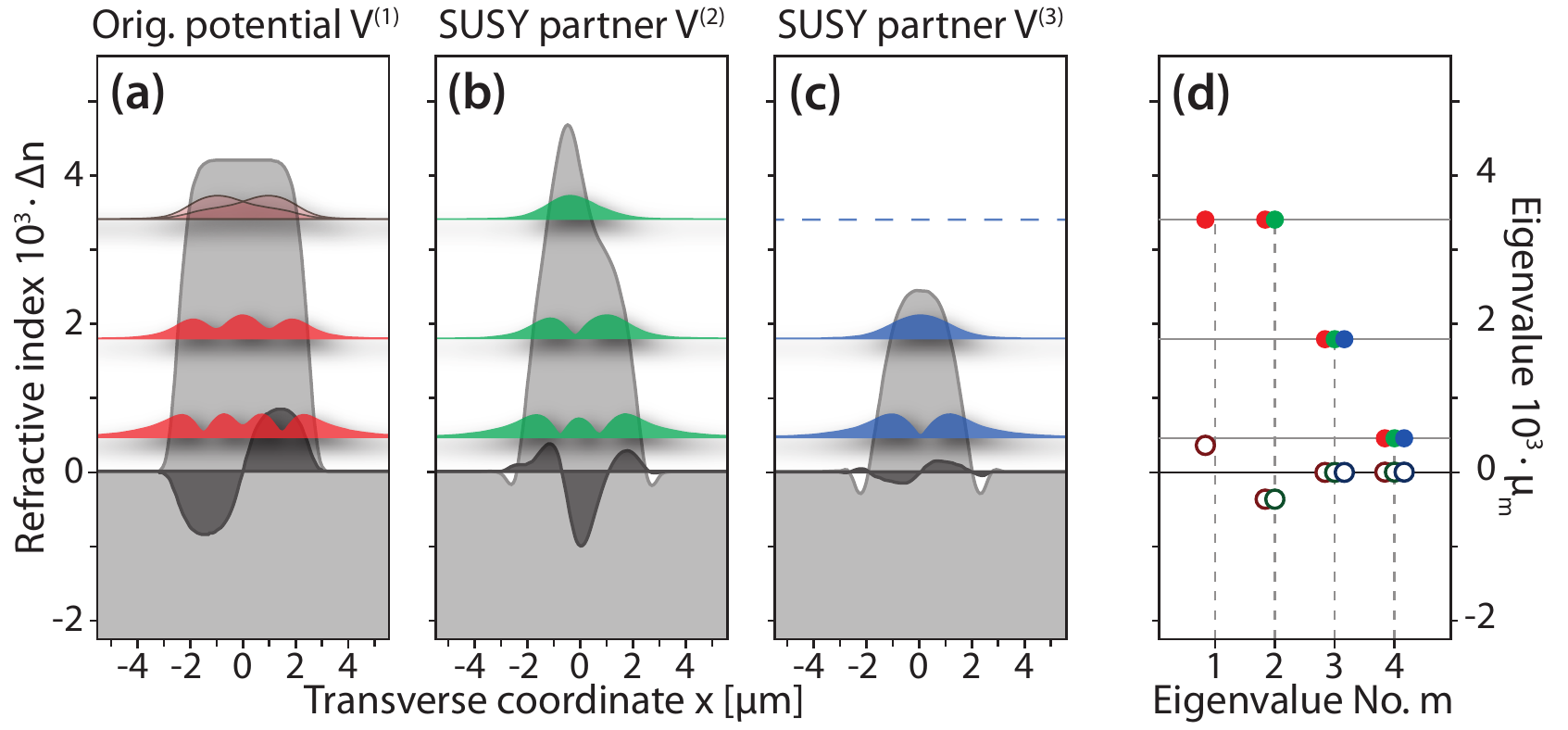}
\caption{(Color online). (a) Refractive index profile of a multimode waveguide supporting a total of four bound states. Here, the imaginary contrast was increased to $\gamma=0.2$ to induce a spontaneous $\mathcal{PT}$ symmetry breaking in the two lowest modes. Sequentially removing the lossy (b) and the amplified (c) mode by means of SUSY restores $\mathcal{PT}$ symmetry in the resulting structure (d).}
\label{fig4}
\end{center}
\end{figure}

In this section we investigate SUSY in systems with spontaneously broken $\mathcal{PT}$ symmetry. When the contrast between gain and loss exceeds a certain threshold, a given real refractive index profile can no longer maintain the symmetry of the bound states. For the waveguide profile of Eq.~\eqref{eq10}, an imaginary contrast of $\gamma=0.2$ places the system well inside this broken-symmetry regime (see Fig.~\ref{fig4}(a)). As it is expected for this type of complex potential \cite{Miri2012_a}, the eigenvalues of the lowest two modes are transformed into a complex conjugate pair with identical real values $\operatorname{Re}\left(\mu_1^{(1)}\right)=\operatorname{Re}\left(\mu_2^{(1)}\right)$ and opposite imaginary parts $\operatorname{Im}\left(\mu_1^{(1)}\right)=-\operatorname{Im}\left(\mu_2^{(1)}\right)$. The corresponding states reside predominantly in the gain (loss) region. Note that the remaining higher order modes retain their $\mathcal{PT}$ symmetry, and therefore continue to exhibit real spectra.

Following the formalism outlined above, SUSY now allows one to remove any of the modes with broken $\mathcal{PT}$ symmetry (Fig.~\ref{fig4}(b)). As in the case of unbroken $\mathcal{PT}$ symmetry, SUSY preserves the remaining set of eigenvalues. In our example, we construct $W$ from the lossy mode ($\operatorname{Im}\left(\mu_1^{(1)}\right)>0$) and hence the partner waveguide supports two neutral modes as well as the remaining mode that in this case experiences amplification. Alternatively, one could have maintained the lossy mode of the system by removing the one subjected to gain ($\operatorname{Im}\left(\mu_2^{(1)}\right)<0$) instead. If both broken-symmetry modes are removed via sequential SUSY transformations (Eqs.~(8,9)), the resulting complex potential $V^{(3)}$ again exhibits an entirely real spectrum and the $\mathcal{PT}$ symmetry of the original structure $V^{(1)}$ is restored (Fig.~\ref{fig4}(c)). The resulting waveguide remains perfectly phase matched to the two neutral modes of the original system (Fig.~\ref{fig4}(d)).

\section{CONCLUSION}

In this work we have shown that the interplay between supersymmetry and $\mathcal{PT}$ symmetry can be fruitfully applied to modify the guided-mode spectra of optical waveguides. In the case of $\mathcal{PT}$-symmetric systems, it preserves the mutual cancelation of gain and loss while allowing for the selective removal of arbitrary guided modes. The SUSY formalism gives rise to isospectral families of complex refractive index landscapes that exhibit entirely real spectra, despite the fact that their shapes violate $\mathcal{PT}$ symmetry. Finally, we have shown how SUSY can facilitate the elimination of modes exhibiting complex eigenvalues in order to overcome the spontaneous breaking of $\mathcal{PT}$ symmetry.

\begin{acknowledgments}

We acknowledge financial support from NSF (grant ECCS-1128520) and AFOSR (grant FA95501210148). M.H. was supported by the German National Academy of Sciences Leopoldina (grant LPDS 2012-01).

\end{acknowledgments}
\bibliography{GF}

\begin{thebibliography}{59}%
\makeatletter
\providecommand \@ifxundefined [1]{%
 \@ifx{#1\undefined}
}%
\providecommand \@ifnum [1]{%
 \ifnum #1\expandafter \@firstoftwo
 \else \expandafter \@secondoftwo
 \fi
}%
\providecommand \@ifx [1]{%
 \ifx #1\expandafter \@firstoftwo
 \else \expandafter \@secondoftwo
 \fi
}%
\providecommand \natexlab [1]{#1}%
\providecommand \enquote  [1]{``#1''}%
\providecommand \bibnamefont  [1]{#1}%
\providecommand \bibfnamefont [1]{#1}%
\providecommand \citenamefont [1]{#1}%
\providecommand \href@noop [0]{\@secondoftwo}%
\providecommand \href [0]{\begingroup \@sanitize@url \@href}%
\providecommand \@href[1]{\@@startlink{#1}\@@href}%
\providecommand \@@href[1]{\endgroup#1\@@endlink}%
\providecommand \@sanitize@url [0]{\catcode `\\12\catcode `\$12\catcode
  `\&12\catcode `\#12\catcode `\^12\catcode `\_12\catcode `\%12\relax}%
\providecommand \@@startlink[1]{}%
\providecommand \@@endlink[0]{}%
\providecommand \url  [0]{\begingroup\@sanitize@url \@url }%
\providecommand \@url [1]{\endgroup\@href {#1}{\urlprefix }}%
\providecommand \urlprefix  [0]{URL }%
\providecommand \Eprint [0]{\href }%
\providecommand \doibase [0]{http://dx.doi.org/}%
\providecommand \selectlanguage [0]{\@gobble}%
\providecommand \bibinfo  [0]{\@secondoftwo}%
\providecommand \bibfield  [0]{\@secondoftwo}%
\providecommand \translation [1]{[#1]}%
\providecommand \BibitemOpen [0]{}%
\providecommand \bibitemStop [0]{}%
\providecommand \bibitemNoStop [0]{.\EOS\space}%
\providecommand \EOS [0]{\spacefactor3000\relax}%
\providecommand \BibitemShut  [1]{\csname bibitem#1\endcsname}%
\let\auto@bib@innerbib\@empty
\bibitem [{\citenamefont {Ramond}(1971)}]{Ram1971}%
  \BibitemOpen
  \bibfield  {author} {\bibinfo {author} {\bibfnamefont {P.}~\bibnamefont
  {Ramond}},\ }\href {\doibase 10.1103/PhysRevD.3.2415} {\bibfield  {journal}
  {\bibinfo  {journal} {Phys. Rev. D}\ }\textbf {\bibinfo {volume} {3}},\
  \bibinfo {pages} {2415} (\bibinfo {year} {1971})}\BibitemShut {NoStop}%
\bibitem [{\citenamefont {Neveu}\ and\ \citenamefont
  {Schwarz}(1971)}]{Nev1971}%
  \BibitemOpen
  \bibfield  {author} {\bibinfo {author} {\bibfnamefont {A.}~\bibnamefont
  {Neveu}}\ and\ \bibinfo {author} {\bibfnamefont {J.~H.}\ \bibnamefont
  {Schwarz}},\ }\href@noop {} {\bibfield  {journal} {\bibinfo  {journal} {Nucl.
  Phys. B}\ }\textbf {\bibinfo {volume} {31}},\ \bibinfo {pages} {86} (\bibinfo
  {year} {1971})}\BibitemShut {NoStop}%
\bibitem [{\citenamefont {Gel'fand}\ and\ \citenamefont
  {Likhtman}(1971)}]{Gel1971}%
  \BibitemOpen
  \bibfield  {author} {\bibinfo {author} {\bibfnamefont {Y.~A.}\ \bibnamefont
  {Gel'fand}}\ and\ \bibinfo {author} {\bibfnamefont {E.~P.}\ \bibnamefont
  {Likhtman}},\ }\href@noop {} {\bibfield  {journal} {\bibinfo  {journal} {JETP
  Lett}\ }\textbf {\bibinfo {volume} {13}},\ \bibinfo {pages} {323} (\bibinfo
  {year} {1971})}\BibitemShut {NoStop}%
\bibitem [{\citenamefont {Volkov}\ and\ \citenamefont
  {Akulov}(1973)}]{Vol1973}%
  \BibitemOpen
  \bibfield  {author} {\bibinfo {author} {\bibfnamefont {D.~V.}\ \bibnamefont
  {Volkov}}\ and\ \bibinfo {author} {\bibfnamefont {V.~P.}\ \bibnamefont
  {Akulov}},\ }\href@noop {} {\bibfield  {journal} {\bibinfo  {journal} {Phys.
  Lett. B}\ }\textbf {\bibinfo {volume} {46}},\ \bibinfo {pages} {109}
  (\bibinfo {year} {1973})}\BibitemShut {NoStop}%
\bibitem [{\citenamefont {Wess}\ and\ \citenamefont {Zumino}(1974)}]{Wes1974}%
  \BibitemOpen
  \bibfield  {author} {\bibinfo {author} {\bibfnamefont {J.}~\bibnamefont
  {Wess}}\ and\ \bibinfo {author} {\bibfnamefont {B.}~\bibnamefont {Zumino}},\
  }\href@noop {} {\bibfield  {journal} {\bibinfo  {journal} {Nucl. Phys. B}\
  }\textbf {\bibinfo {volume} {70}},\ \bibinfo {pages} {39} (\bibinfo {year}
  {1974})}\BibitemShut {NoStop}%
\bibitem [{\citenamefont {Witten}(1981)}]{Wit1981}%
  \BibitemOpen
  \bibfield  {author} {\bibinfo {author} {\bibfnamefont {E.}~\bibnamefont
  {Witten}},\ }\href@noop {} {\bibfield  {journal} {\bibinfo  {journal} {Nucl.
  Phys. B}\ }\textbf {\bibinfo {volume} {185}},\ \bibinfo {pages} {513}
  (\bibinfo {year} {1981})}\BibitemShut {NoStop}%
\bibitem [{\citenamefont {Cooper}\ and\ \citenamefont
  {Freedman}(1983)}]{Coo1983}%
  \BibitemOpen
  \bibfield  {author} {\bibinfo {author} {\bibfnamefont {F.}~\bibnamefont
  {Cooper}}\ and\ \bibinfo {author} {\bibfnamefont {B.}~\bibnamefont
  {Freedman}},\ }\href@noop {} {\bibfield  {journal} {\bibinfo  {journal} {Ann.
  Phys.}\ }\textbf {\bibinfo {volume} {146}},\ \bibinfo {pages} {262} (\bibinfo
  {year} {1983})}\BibitemShut {NoStop}%
\bibitem [{\citenamefont {Blockley}\ and\ \citenamefont
  {Stedman}(1985{\natexlab{a}})}]{Ste1985_1}%
  \BibitemOpen
  \bibfield  {author} {\bibinfo {author} {\bibfnamefont {C.~A.}\ \bibnamefont
  {Blockley}}\ and\ \bibinfo {author} {\bibfnamefont {G.~A.}\ \bibnamefont
  {Stedman}},\ }\href@noop {} {\bibfield  {journal} {\bibinfo  {journal} {Eur.
  J. Phys.}\ }\textbf {\bibinfo {volume} {6}},\ \bibinfo {pages} {218}
  (\bibinfo {year} {1985}{\natexlab{a}})}\BibitemShut {NoStop}%
\bibitem [{\citenamefont {Blockley}\ and\ \citenamefont
  {Stedman}(1985{\natexlab{b}})}]{Ste1985_2}%
  \BibitemOpen
  \bibfield  {author} {\bibinfo {author} {\bibfnamefont {C.~A.}\ \bibnamefont
  {Blockley}}\ and\ \bibinfo {author} {\bibfnamefont {G.~A.}\ \bibnamefont
  {Stedman}},\ }\href@noop {} {\bibfield  {journal} {\bibinfo  {journal} {Eur.
  J. Phys.}\ }\textbf {\bibinfo {volume} {6}},\ \bibinfo {pages} {225}
  (\bibinfo {year} {1985}{\natexlab{b}})}\BibitemShut {NoStop}%
\bibitem [{\citenamefont {Khare}\ and\ \citenamefont
  {Sukhatme}(1989)}]{Kha1989}%
  \BibitemOpen
  \bibfield  {author} {\bibinfo {author} {\bibfnamefont {A.}~\bibnamefont
  {Khare}}\ and\ \bibinfo {author} {\bibfnamefont {U.}~\bibnamefont
  {Sukhatme}},\ }\href@noop {} {\bibfield  {journal} {\bibinfo  {journal} {J.
  Phys. A: Math. Gen.}\ }\textbf {\bibinfo {volume} {22}},\ \bibinfo {pages}
  {2847} (\bibinfo {year} {1989})}\BibitemShut {NoStop}%
\bibitem [{\citenamefont {Gendenshtein}\ and\ \citenamefont
  {Krive}(1985)}]{Gen1985}%
  \BibitemOpen
  \bibfield  {author} {\bibinfo {author} {\bibfnamefont {L.~E.}\ \bibnamefont
  {Gendenshtein}}\ and\ \bibinfo {author} {\bibfnamefont {I.~V.}\ \bibnamefont
  {Krive}},\ }\href@noop {} {\bibfield  {journal} {\bibinfo  {journal} {Usp.
  Fiz. Nauk}\ }\textbf {\bibinfo {volume} {146}},\ \bibinfo {pages} {553}
  (\bibinfo {year} {1985})}\BibitemShut {NoStop}%
\bibitem [{\citenamefont {Lahiri}\ \emph {et~al.}(1990)\citenamefont {Lahiri},
  \citenamefont {Roy},\ and\ \citenamefont {Bagchi}}]{Lah1990}%
  \BibitemOpen
  \bibfield  {author} {\bibinfo {author} {\bibfnamefont {A.}~\bibnamefont
  {Lahiri}}, \bibinfo {author} {\bibfnamefont {P.}~\bibnamefont {Roy}}, \ and\
  \bibinfo {author} {\bibfnamefont {B.}~\bibnamefont {Bagchi}},\ }\href@noop {}
  {\bibfield  {journal} {\bibinfo  {journal} {Int. Jour. Mod. Phys. A}\
  }\textbf {\bibinfo {volume} {5}},\ \bibinfo {pages} {1383} (\bibinfo {year}
  {1990})}\BibitemShut {NoStop}%
\bibitem [{\citenamefont {Cooper}\ \emph {et~al.}(1995)\citenamefont {Cooper},
  \citenamefont {Khare},\ and\ \citenamefont {Sukhatme}}]{Coo1995}%
  \BibitemOpen
  \bibfield  {author} {\bibinfo {author} {\bibfnamefont {F.}~\bibnamefont
  {Cooper}}, \bibinfo {author} {\bibfnamefont {A.}~\bibnamefont {Khare}}, \
  and\ \bibinfo {author} {\bibfnamefont {U.}~\bibnamefont {Sukhatme}},\
  }\href@noop {} {\bibfield  {journal} {\bibinfo  {journal} {Phys. Rep.}\
  }\textbf {\bibinfo {volume} {251}},\ \bibinfo {pages} {267} (\bibinfo {year}
  {1995})}\BibitemShut {NoStop}%
\bibitem [{\citenamefont {{Miri}}\ \emph {et~al.}(2013)\citenamefont {{Miri}},
  \citenamefont {{Heinrich}}, \citenamefont {{El-Ganainy}},\ and\ \citenamefont
  {{Christodoulides}}}]{Miri2013}%
  \BibitemOpen
  \bibfield  {author} {\bibinfo {author} {\bibfnamefont {M.-A.}\ \bibnamefont
  {{Miri}}}, \bibinfo {author} {\bibfnamefont {M.}~\bibnamefont {{Heinrich}}},
  \bibinfo {author} {\bibfnamefont {R.}~\bibnamefont {{El-Ganainy}}}, \ and\
  \bibinfo {author} {\bibfnamefont {D.~N.}\ \bibnamefont {{Christodoulides}}},\
  }\href@noop {} {\bibfield  {journal} {\bibinfo  {journal} {ArXiv e-prints}\ }
  (\bibinfo {year} {2013})},\ \Eprint {http://arxiv.org/abs/1304.6646}
  {arXiv:1304.6646 [physics.optics]} \BibitemShut {NoStop}%
\bibitem [{\citenamefont {Bender}\ and\ \citenamefont
  {Boettcher}(1998)}]{Ben1998}%
  \BibitemOpen
  \bibfield  {author} {\bibinfo {author} {\bibfnamefont {C.~M.}\ \bibnamefont
  {Bender}}\ and\ \bibinfo {author} {\bibfnamefont {S.}~\bibnamefont
  {Boettcher}},\ }\href {\doibase 10.1103/PhysRevLett.80.5243} {\bibfield
  {journal} {\bibinfo  {journal} {Phys. Rev. Lett.}\ }\textbf {\bibinfo
  {volume} {80}},\ \bibinfo {pages} {5243} (\bibinfo {year}
  {1998})}\BibitemShut {NoStop}%
\bibitem [{\citenamefont {Bender}\ \emph {et~al.}(2002)\citenamefont {Bender},
  \citenamefont {Brody},\ and\ \citenamefont {Jones}}]{Ben2002}%
  \BibitemOpen
  \bibfield  {author} {\bibinfo {author} {\bibfnamefont {C.~M.}\ \bibnamefont
  {Bender}}, \bibinfo {author} {\bibfnamefont {D.~C.}\ \bibnamefont {Brody}}, \
  and\ \bibinfo {author} {\bibfnamefont {H.~F.}\ \bibnamefont {Jones}},\ }\href
  {\doibase 10.1103/PhysRevLett.89.270401} {\bibfield  {journal} {\bibinfo
  {journal} {Phys. Rev. Lett.}\ }\textbf {\bibinfo {volume} {89}},\ \bibinfo
  {pages} {270401} (\bibinfo {year} {2002})}\BibitemShut {NoStop}%
\bibitem [{\citenamefont {Bender}(2007)}]{Ben2007}%
  \BibitemOpen
  \bibfield  {author} {\bibinfo {author} {\bibfnamefont {C.~M.}\ \bibnamefont
  {Bender}},\ }\href {\doibase 10.1088/0034-4885/70/6/R03} {\bibfield
  {journal} {\bibinfo  {journal} {Rep. Prog. Phys.}\ }\textbf {\bibinfo
  {volume} {70}},\ \bibinfo {pages} {947} (\bibinfo {year} {2007})}\BibitemShut
  {NoStop}%
\bibitem [{\citenamefont {Lévai}\ and\ \citenamefont {Znojil}(2000)}]{Lev2001}%
  \BibitemOpen
  \bibfield  {author} {\bibinfo {author} {\bibfnamefont {G.}~\bibnamefont
  {Lévai}}\ and\ \bibinfo {author} {\bibfnamefont {M.}~\bibnamefont {Znojil}},\
  }\href {\doibase 10.1088/0305-4470/33/40/313} {\bibfield  {journal} {\bibinfo
   {journal} {J. Phys. A: Math. Gen.}\ }\textbf {\bibinfo {volume} {33}},\
  \bibinfo {pages} {7165} (\bibinfo {year} {2000})}\BibitemShut {NoStop}%
\bibitem [{\citenamefont {Ahmed}(2001)}]{Ahm2001}%
  \BibitemOpen
  \bibfield  {author} {\bibinfo {author} {\bibfnamefont {Z.}~\bibnamefont
  {Ahmed}},\ }\href {\doibase 10.1016/S0375-9601(01)00218-3} {\bibfield
  {journal} {\bibinfo  {journal} {Phys. Lett. A}\ }\textbf {\bibinfo {volume}
  {282}},\ \bibinfo {pages} {343} (\bibinfo {year} {2001})}\BibitemShut
  {NoStop}%
\bibitem [{\citenamefont {Ahmed}\ \emph {et~al.}(2005)\citenamefont {Ahmed},
  \citenamefont {Bender},\ and\ \citenamefont {Berry}}]{Ahm2005}%
  \BibitemOpen
  \bibfield  {author} {\bibinfo {author} {\bibfnamefont {Z.}~\bibnamefont
  {Ahmed}}, \bibinfo {author} {\bibfnamefont {C.~M.}\ \bibnamefont {Bender}}, \
  and\ \bibinfo {author} {\bibfnamefont {M.~V.}\ \bibnamefont {Berry}},\ }\href
  {\doibase 10.1088/0305-4470/38/39/L01} {\bibfield  {journal} {\bibinfo
  {journal} {J. Phys. A: Math. Gen.}\ }\textbf {\bibinfo {volume} {38}},\
  \bibinfo {pages} {L627} (\bibinfo {year} {2005})}\BibitemShut {NoStop}%
\bibitem [{\citenamefont {Mostafazadeh}(2002{\natexlab{a}})}]{Mos2002_a}%
  \BibitemOpen
  \bibfield  {author} {\bibinfo {author} {\bibfnamefont {A.}~\bibnamefont
  {Mostafazadeh}},\ }\href {\doibase 10.1063/1.1418246} {\bibfield  {journal}
  {\bibinfo  {journal} {J. Math. Phys.}\ }\textbf {\bibinfo {volume} {43}},\
  \bibinfo {pages} {205} (\bibinfo {year} {2002}{\natexlab{a}})}\BibitemShut
  {NoStop}%
\bibitem [{\citenamefont {Mostafazadeh}(2002{\natexlab{b}})}]{Mos2002_b}%
  \BibitemOpen
  \bibfield  {author} {\bibinfo {author} {\bibfnamefont {A.}~\bibnamefont
  {Mostafazadeh}},\ }\href {\doibase 10.1063/1.1461427} {\bibfield  {journal}
  {\bibinfo  {journal} {J. Math. Phys.}\ }\textbf {\bibinfo {volume} {43}},\
  \bibinfo {pages} {2814} (\bibinfo {year} {2002}{\natexlab{b}})}\BibitemShut
  {NoStop}%
\bibitem [{\citenamefont {Cannata}\ \emph {et~al.}(1998)\citenamefont
  {Cannata}, \citenamefont {Junker},\ and\ \citenamefont {Trost}}]{Can1998}%
  \BibitemOpen
  \bibfield  {author} {\bibinfo {author} {\bibfnamefont {F.}~\bibnamefont
  {Cannata}}, \bibinfo {author} {\bibfnamefont {G.}~\bibnamefont {Junker}}, \
  and\ \bibinfo {author} {\bibfnamefont {J.}~\bibnamefont {Trost}},\ }\href
  {\doibase 10.1016/S0375-9601(98)00517-9} {\bibfield  {journal} {\bibinfo
  {journal} {Phys. Lett. A.}\ }\textbf {\bibinfo {volume} {246}},\ \bibinfo
  {pages} {219} (\bibinfo {year} {1998})}\BibitemShut {NoStop}%
\bibitem [{\citenamefont {Andrianov}\ \emph {et~al.}(1999)\citenamefont
  {Andrianov}, \citenamefont {Ioffe}, \citenamefont {Cannata},\ and\
  \citenamefont {Dedonder}}]{And1999}%
  \BibitemOpen
  \bibfield  {author} {\bibinfo {author} {\bibfnamefont {A.~A.}\ \bibnamefont
  {Andrianov}}, \bibinfo {author} {\bibfnamefont {M.~V.}\ \bibnamefont
  {Ioffe}}, \bibinfo {author} {\bibfnamefont {F.}~\bibnamefont {Cannata}}, \
  and\ \bibinfo {author} {\bibfnamefont {J.-P.}\ \bibnamefont {Dedonder}},\
  }\href {\doibase 10.1142/S0217751X99001342} {\bibfield  {journal} {\bibinfo
  {journal} {Int. J. Mod. Phys. A}\ }\textbf {\bibinfo {volume} {14}},\
  \bibinfo {pages} {2675} (\bibinfo {year} {1999})}\BibitemShut {NoStop}%
\bibitem [{\citenamefont {Bagchi}\ \emph {et~al.}(2000)\citenamefont {Bagchi},
  \citenamefont {Cannata},\ and\ \citenamefont {Quesne}}]{Bag2000}%
  \BibitemOpen
  \bibfield  {author} {\bibinfo {author} {\bibfnamefont {B.}~\bibnamefont
  {Bagchi}}, \bibinfo {author} {\bibfnamefont {F.}~\bibnamefont {Cannata}}, \
  and\ \bibinfo {author} {\bibfnamefont {C.}~\bibnamefont {Quesne}},\ }\href
  {\doibase 10.1016/S0375-9601(00)00227-9} {\bibfield  {journal} {\bibinfo
  {journal} {Phys. Lett. A.}\ }\textbf {\bibinfo {volume} {269}},\ \bibinfo
  {pages} {79} (\bibinfo {year} {2000})}\BibitemShut {NoStop}%
\bibitem [{\citenamefont {Znojil}\ \emph {et~al.}(2000)\citenamefont {Znojil},
  \citenamefont {Cannata}, \citenamefont {Bagchi},\ and\ \citenamefont
  {Roychoudhury}}]{Zno2000}%
  \BibitemOpen
  \bibfield  {author} {\bibinfo {author} {\bibfnamefont {M.}~\bibnamefont
  {Znojil}}, \bibinfo {author} {\bibfnamefont {F.}~\bibnamefont {Cannata}},
  \bibinfo {author} {\bibfnamefont {B.}~\bibnamefont {Bagchi}}, \ and\ \bibinfo
  {author} {\bibfnamefont {R.}~\bibnamefont {Roychoudhury}},\ }\href {\doibase
  10.1016/S0370-2693(00)00569-4} {\bibfield  {journal} {\bibinfo  {journal}
  {Phys. Lett. B}\ }\textbf {\bibinfo {volume} {483}},\ \bibinfo {pages} {284}
  (\bibinfo {year} {2000})}\BibitemShut {NoStop}%
\bibitem [{\citenamefont {Dorey}\ \emph {et~al.}(2001)\citenamefont {Dorey},
  \citenamefont {Dunning},\ and\ \citenamefont {Tateo}}]{Dor2001}%
  \BibitemOpen
  \bibfield  {author} {\bibinfo {author} {\bibfnamefont {P.}~\bibnamefont
  {Dorey}}, \bibinfo {author} {\bibfnamefont {C.}~\bibnamefont {Dunning}}, \
  and\ \bibinfo {author} {\bibfnamefont {R.}~\bibnamefont {Tateo}},\ }\href
  {\doibase 10.1088/0305-4470/34/28/102} {\bibfield  {journal} {\bibinfo
  {journal} {J. Phys. A: Math. Gen.}\ }\textbf {\bibinfo {volume} {34}},\
  \bibinfo {pages} {L391} (\bibinfo {year} {2001})}\BibitemShut {NoStop}%
\bibitem [{\citenamefont {Cannata}\ \emph {et~al.}(2001)\citenamefont
  {Cannata}, \citenamefont {Ioffe}, \citenamefont {Roychoudhury},\ and\
  \citenamefont {Roy}}]{Can2001}%
  \BibitemOpen
  \bibfield  {author} {\bibinfo {author} {\bibfnamefont {F.}~\bibnamefont
  {Cannata}}, \bibinfo {author} {\bibfnamefont {M.}~\bibnamefont {Ioffe}},
  \bibinfo {author} {\bibfnamefont {R.}~\bibnamefont {Roychoudhury}}, \ and\
  \bibinfo {author} {\bibfnamefont {P.}~\bibnamefont {Roy}},\ }\href {\doibase
  10.1016/S0375-9601(01)00144-X} {\bibfield  {journal} {\bibinfo  {journal}
  {Phys. Lett. A}\ }\textbf {\bibinfo {volume} {281}},\ \bibinfo {pages} {305}
  (\bibinfo {year} {2001})}\BibitemShut {NoStop}%
\bibitem [{\citenamefont {Levai}\ and\ \citenamefont {Znojil}(2002)}]{Lev2002}%
  \BibitemOpen
  \bibfield  {author} {\bibinfo {author} {\bibfnamefont {G.}~\bibnamefont
  {Levai}}\ and\ \bibinfo {author} {\bibfnamefont {M.}~\bibnamefont {Znojil}},\
  }\href {\doibase 10.1088/0305-4470/35/41/311} {\bibfield  {journal} {\bibinfo
   {journal} {J. Phys. A: Math. Gen.}\ }\textbf {\bibinfo {volume} {35}},\
  \bibinfo {pages} {8793} (\bibinfo {year} {2002})}\BibitemShut {NoStop}%
\bibitem [{\citenamefont {Bermudez}\ and\ \citenamefont
  {Fernández~C.}(2011)}]{Ber2011}%
  \BibitemOpen
  \bibfield  {author} {\bibinfo {author} {\bibfnamefont {D.}~\bibnamefont
  {Bermudez}}\ and\ \bibinfo {author} {\bibfnamefont {D.~J.}\ \bibnamefont
  {Fernández~C.}},\ }\href {\doibase 10.1016/j.physleta.2011.06.042} {\bibfield
   {journal} {\bibinfo  {journal} {Phys. Lett. A.}\ }\textbf {\bibinfo {volume}
  {375}},\ \bibinfo {pages} {2974} (\bibinfo {year} {2011})}\BibitemShut
  {NoStop}%
\bibitem [{\citenamefont {El-Ganainy}\ \emph {et~al.}(2007)\citenamefont
  {El-Ganainy}, \citenamefont {Makris}, \citenamefont {Christodoulides},\ and\
  \citenamefont {Musslimani}}]{Elg2007}%
  \BibitemOpen
  \bibfield  {author} {\bibinfo {author} {\bibfnamefont {R.}~\bibnamefont
  {El-Ganainy}}, \bibinfo {author} {\bibfnamefont {K.~G.}\ \bibnamefont
  {Makris}}, \bibinfo {author} {\bibfnamefont {D.~N.}\ \bibnamefont
  {Christodoulides}}, \ and\ \bibinfo {author} {\bibfnamefont {Z.~H.}\
  \bibnamefont {Musslimani}},\ }\href {\doibase 10.1364/OL.32.002632}
  {\bibfield  {journal} {\bibinfo  {journal} {Opt. Lett.}\ }\textbf {\bibinfo
  {volume} {32}},\ \bibinfo {pages} {2632} (\bibinfo {year}
  {2007})}\BibitemShut {NoStop}%
\bibitem [{\citenamefont {Makris}\ \emph {et~al.}(2008)\citenamefont {Makris},
  \citenamefont {El-Ganainy}, \citenamefont {Christodoulides},\ and\
  \citenamefont {Musslimani}}]{Mak2008}%
  \BibitemOpen
  \bibfield  {author} {\bibinfo {author} {\bibfnamefont {K.~G.}\ \bibnamefont
  {Makris}}, \bibinfo {author} {\bibfnamefont {R.}~\bibnamefont {El-Ganainy}},
  \bibinfo {author} {\bibfnamefont {D.~N.}\ \bibnamefont {Christodoulides}}, \
  and\ \bibinfo {author} {\bibfnamefont {Z.~H.}\ \bibnamefont {Musslimani}},\
  }\href {\doibase 10.1103/PhysRevLett.100.103904} {\bibfield  {journal}
  {\bibinfo  {journal} {Phys. Rev. Lett.}\ }\textbf {\bibinfo {volume} {100}},\
  \bibinfo {pages} {103904} (\bibinfo {year} {2008})}\BibitemShut {NoStop}%
\bibitem [{\citenamefont {Musslimani}\ \emph {et~al.}(2008)\citenamefont
  {Musslimani}, \citenamefont {Makris}, \citenamefont {El-Ganainy},\ and\
  \citenamefont {Christodoulides}}]{Mus2008}%
  \BibitemOpen
  \bibfield  {author} {\bibinfo {author} {\bibfnamefont {Z.~H.}\ \bibnamefont
  {Musslimani}}, \bibinfo {author} {\bibfnamefont {K.~G.}\ \bibnamefont
  {Makris}}, \bibinfo {author} {\bibfnamefont {R.}~\bibnamefont {El-Ganainy}},
  \ and\ \bibinfo {author} {\bibfnamefont {D.~N.}\ \bibnamefont
  {Christodoulides}},\ }\href {\doibase 10.1103/PhysRevLett.100.030402}
  {\bibfield  {journal} {\bibinfo  {journal} {Phys. Rev. Lett.}\ }\textbf
  {\bibinfo {volume} {100}},\ \bibinfo {pages} {030402} (\bibinfo {year}
  {2008})}\BibitemShut {NoStop}%
\bibitem [{\citenamefont {Regensburger}\ \emph {et~al.}(2012)\citenamefont
  {Regensburger}, \citenamefont {Bersch}, \citenamefont {Miri}, \citenamefont
  {Onishchukov}, \citenamefont {Christodoulides},\ and\ \citenamefont
  {Peschel}}]{Reg2012}%
  \BibitemOpen
  \bibfield  {author} {\bibinfo {author} {\bibfnamefont {A.}~\bibnamefont
  {Regensburger}}, \bibinfo {author} {\bibfnamefont {C.}~\bibnamefont
  {Bersch}}, \bibinfo {author} {\bibfnamefont {M.-A.}\ \bibnamefont {Miri}},
  \bibinfo {author} {\bibfnamefont {G.}~\bibnamefont {Onishchukov}}, \bibinfo
  {author} {\bibfnamefont {D.~N.}\ \bibnamefont {Christodoulides}}, \ and\
  \bibinfo {author} {\bibfnamefont {U.}~\bibnamefont {Peschel}},\ }\href
  {\doibase 10.1038/nature11298} {\ \textbf {\bibinfo {volume} {488}},\
  \bibinfo {pages} {167} (\bibinfo {year} {2012})}\BibitemShut {NoStop}%
\bibitem [{\citenamefont {Ruter}\ \emph {et~al.}(2010)\citenamefont {Ruter},
  \citenamefont {Makris}, \citenamefont {El-Ganainy}, \citenamefont
  {Christodoulides}, \citenamefont {Segev},\ and\ \citenamefont
  {Kip}}]{Rut2010}%
  \BibitemOpen
  \bibfield  {author} {\bibinfo {author} {\bibfnamefont {C.~E.}\ \bibnamefont
  {Ruter}}, \bibinfo {author} {\bibfnamefont {K.~G.}\ \bibnamefont {Makris}},
  \bibinfo {author} {\bibfnamefont {R.}~\bibnamefont {El-Ganainy}}, \bibinfo
  {author} {\bibfnamefont {D.~N.}\ \bibnamefont {Christodoulides}}, \bibinfo
  {author} {\bibfnamefont {M.}~\bibnamefont {Segev}}, \ and\ \bibinfo {author}
  {\bibfnamefont {D.}~\bibnamefont {Kip}},\ }\href {\doibase 10.1038/nphys1515}
  {\bibfield  {journal} {\bibinfo  {journal} {Nature. Phys.}\ }\textbf
  {\bibinfo {volume} {6}},\ \bibinfo {pages} {192} (\bibinfo {year}
  {2010})}\BibitemShut {NoStop}%
\bibitem [{\citenamefont {Guo}\ \emph {et~al.}(2009)\citenamefont {Guo},
  \citenamefont {Salamo}, \citenamefont {Duchesne}, \citenamefont {Morandotti},
  \citenamefont {Volatier-Ravat}, \citenamefont {Aimez}, \citenamefont
  {Siviloglou},\ and\ \citenamefont {Christodoulides}}]{Guo2010}%
  \BibitemOpen
  \bibfield  {author} {\bibinfo {author} {\bibfnamefont {A.}~\bibnamefont
  {Guo}}, \bibinfo {author} {\bibfnamefont {G.~J.}\ \bibnamefont {Salamo}},
  \bibinfo {author} {\bibfnamefont {D.}~\bibnamefont {Duchesne}}, \bibinfo
  {author} {\bibfnamefont {R.}~\bibnamefont {Morandotti}}, \bibinfo {author}
  {\bibfnamefont {M.}~\bibnamefont {Volatier-Ravat}}, \bibinfo {author}
  {\bibfnamefont {V.}~\bibnamefont {Aimez}}, \bibinfo {author} {\bibfnamefont
  {G.~A.}\ \bibnamefont {Siviloglou}}, \ and\ \bibinfo {author} {\bibfnamefont
  {D.~N.}\ \bibnamefont {Christodoulides}},\ }\href {\doibase
  10.1103/PhysRevLett.103.093902} {\bibfield  {journal} {\bibinfo  {journal}
  {Phys. Rev. Lett.}\ }\textbf {\bibinfo {volume} {103}},\ \bibinfo {pages}
  {093902} (\bibinfo {year} {2009})}\BibitemShut {NoStop}%
\bibitem [{\citenamefont {Feng}\ \emph {et~al.}(2013)\citenamefont {Feng},
  \citenamefont {Xu}, \citenamefont {Fegadolli}, \citenamefont {Lu},
  \citenamefont {Oliveira}, \citenamefont {Almeida}, \citenamefont {Chen},\
  and\ \citenamefont {Scherer}}]{Fen2013}%
  \BibitemOpen
  \bibfield  {author} {\bibinfo {author} {\bibfnamefont {L.}~\bibnamefont
  {Feng}}, \bibinfo {author} {\bibfnamefont {Y.-L.}\ \bibnamefont {Xu}},
  \bibinfo {author} {\bibfnamefont {W.~S.}\ \bibnamefont {Fegadolli}}, \bibinfo
  {author} {\bibfnamefont {M.-H.}\ \bibnamefont {Lu}}, \bibinfo {author}
  {\bibfnamefont {J.~E.~B.}\ \bibnamefont {Oliveira}}, \bibinfo {author}
  {\bibfnamefont {V.~R.}\ \bibnamefont {Almeida}}, \bibinfo {author}
  {\bibfnamefont {Y.-F.}\ \bibnamefont {Chen}}, \ and\ \bibinfo {author}
  {\bibfnamefont {A.}~\bibnamefont {Scherer}},\ }\href {\doibase
  10.1038/nmat3495} {\bibfield  {journal} {\bibinfo  {journal} {Nature. Matt.}\
  }\textbf {\bibinfo {volume} {12}},\ \bibinfo {pages} {108} (\bibinfo {year}
  {2013})}\BibitemShut {NoStop}%
\bibitem [{\citenamefont {Klaiman}\ \emph {et~al.}(2008)\citenamefont
  {Klaiman}, \citenamefont {G\"unther},\ and\ \citenamefont
  {Moiseyev}}]{Kla2008}%
  \BibitemOpen
  \bibfield  {author} {\bibinfo {author} {\bibfnamefont {S.}~\bibnamefont
  {Klaiman}}, \bibinfo {author} {\bibfnamefont {U.}~\bibnamefont {G\"unther}},
  \ and\ \bibinfo {author} {\bibfnamefont {N.}~\bibnamefont {Moiseyev}},\
  }\href {\doibase 10.1103/PhysRevLett.101.080402} {\bibfield  {journal}
  {\bibinfo  {journal} {Phys. Rev. Lett.}\ }\textbf {\bibinfo {volume} {101}},\
  \bibinfo {pages} {080402} (\bibinfo {year} {2008})}\BibitemShut {NoStop}%
\bibitem [{\citenamefont {Longhi}(2009)}]{Lon2009}%
  \BibitemOpen
  \bibfield  {author} {\bibinfo {author} {\bibfnamefont {S.}~\bibnamefont
  {Longhi}},\ }\href {\doibase 10.1103/PhysRevLett.103.123601} {\bibfield
  {journal} {\bibinfo  {journal} {Phys. Rev. Lett.}\ }\textbf {\bibinfo
  {volume} {103}},\ \bibinfo {pages} {123601} (\bibinfo {year}
  {2009})}\BibitemShut {NoStop}%
\bibitem [{\citenamefont {Longhi}(2010)}]{Lon2010}%
  \BibitemOpen
  \bibfield  {author} {\bibinfo {author} {\bibfnamefont {S.}~\bibnamefont
  {Longhi}},\ }\href {\doibase 10.1103/PhysRevA.82.031801} {\bibfield
  {journal} {\bibinfo  {journal} {Phys. Rev. A}\ }\textbf {\bibinfo {volume}
  {82}},\ \bibinfo {pages} {031801} (\bibinfo {year} {2010})}\BibitemShut
  {NoStop}%
\bibitem [{\citenamefont {Schomerus}(2010)}]{Sch2010}%
  \BibitemOpen
  \bibfield  {author} {\bibinfo {author} {\bibfnamefont {H.}~\bibnamefont
  {Schomerus}},\ }\href {\doibase 10.1103/PhysRevLett.104.233601} {\bibfield
  {journal} {\bibinfo  {journal} {Phys. Rev. Lett.}\ }\textbf {\bibinfo
  {volume} {104}},\ \bibinfo {pages} {233601} (\bibinfo {year}
  {2010})}\BibitemShut {NoStop}%
\bibitem [{\citenamefont {Lin}\ \emph {et~al.}(2011)\citenamefont {Lin},
  \citenamefont {Ramezani}, \citenamefont {Eichelkraut}, \citenamefont
  {Kottos}, \citenamefont {Cao},\ and\ \citenamefont
  {Christodoulides}}]{Lin2011}%
  \BibitemOpen
  \bibfield  {author} {\bibinfo {author} {\bibfnamefont {Z.}~\bibnamefont
  {Lin}}, \bibinfo {author} {\bibfnamefont {H.}~\bibnamefont {Ramezani}},
  \bibinfo {author} {\bibfnamefont {T.}~\bibnamefont {Eichelkraut}}, \bibinfo
  {author} {\bibfnamefont {T.}~\bibnamefont {Kottos}}, \bibinfo {author}
  {\bibfnamefont {H.}~\bibnamefont {Cao}}, \ and\ \bibinfo {author}
  {\bibfnamefont {D.~N.}\ \bibnamefont {Christodoulides}},\ }\href {\doibase
  10.1103/PhysRevLett.106.213901} {\bibfield  {journal} {\bibinfo  {journal}
  {Phys. Rev. Lett.}\ }\textbf {\bibinfo {volume} {106}},\ \bibinfo {pages}
  {213901} (\bibinfo {year} {2011})}\BibitemShut {NoStop}%
\bibitem [{\citenamefont {Kulishov}\ \emph {et~al.}(2005)\citenamefont
  {Kulishov}, \citenamefont {Laniel}, \citenamefont {Belanger}, \citenamefont
  {Azana},\ and\ \citenamefont {Plant}}]{Kul2005}%
  \BibitemOpen
  \bibfield  {author} {\bibinfo {author} {\bibfnamefont {M.}~\bibnamefont
  {Kulishov}}, \bibinfo {author} {\bibfnamefont {J.}~\bibnamefont {Laniel}},
  \bibinfo {author} {\bibfnamefont {N.}~\bibnamefont {Belanger}}, \bibinfo
  {author} {\bibfnamefont {J.}~\bibnamefont {Azana}}, \ and\ \bibinfo {author}
  {\bibfnamefont {D.}~\bibnamefont {Plant}},\ }\href@noop {} {\bibfield
  {journal} {\bibinfo  {journal} {Opt. Express.}\ }\textbf {\bibinfo {volume}
  {13}},\ \bibinfo {pages} {3068} (\bibinfo {year} {2005})}\BibitemShut
  {NoStop}%
\bibitem [{\citenamefont {Sukhorukov}\ \emph {et~al.}(2010)\citenamefont
  {Sukhorukov}, \citenamefont {Xu},\ and\ \citenamefont {Kivshar}}]{Suk2010}%
  \BibitemOpen
  \bibfield  {author} {\bibinfo {author} {\bibfnamefont {A.~A.}\ \bibnamefont
  {Sukhorukov}}, \bibinfo {author} {\bibfnamefont {Z.}~\bibnamefont {Xu}}, \
  and\ \bibinfo {author} {\bibfnamefont {Y.~S.}\ \bibnamefont {Kivshar}},\
  }\href {\doibase 10.1103/PhysRevA.82.043818} {\bibfield  {journal} {\bibinfo
  {journal} {Phys. Rev. A}\ }\textbf {\bibinfo {volume} {82}},\ \bibinfo
  {pages} {043818} (\bibinfo {year} {2010})}\BibitemShut {NoStop}%
\bibitem [{\citenamefont {Miroshnichenko}\ \emph {et~al.}(2011)\citenamefont
  {Miroshnichenko}, \citenamefont {Malomed},\ and\ \citenamefont
  {Kivshar}}]{Mir2011}%
  \BibitemOpen
  \bibfield  {author} {\bibinfo {author} {\bibfnamefont {A.~E.}\ \bibnamefont
  {Miroshnichenko}}, \bibinfo {author} {\bibfnamefont {B.~A.}\ \bibnamefont
  {Malomed}}, \ and\ \bibinfo {author} {\bibfnamefont {Y.~S.}\ \bibnamefont
  {Kivshar}},\ }\href {\doibase 10.1103/PhysRevA.84.012123} {\bibfield
  {journal} {\bibinfo  {journal} {Phys. Rev. A}\ }\textbf {\bibinfo {volume}
  {84}},\ \bibinfo {pages} {012123} (\bibinfo {year} {2011})}\BibitemShut
  {NoStop}%
\bibitem [{\citenamefont {Chong}\ \emph {et~al.}(2011)\citenamefont {Chong},
  \citenamefont {Ge},\ and\ \citenamefont {Stone}}]{Cho2011}%
  \BibitemOpen
  \bibfield  {author} {\bibinfo {author} {\bibfnamefont {Y.~D.}\ \bibnamefont
  {Chong}}, \bibinfo {author} {\bibfnamefont {L.}~\bibnamefont {Ge}}, \ and\
  \bibinfo {author} {\bibfnamefont {A.~D.}\ \bibnamefont {Stone}},\ }\href
  {\doibase 10.1103/PhysRevLett.106.093902} {\bibfield  {journal} {\bibinfo
  {journal} {Phys. Rev. Lett.}\ }\textbf {\bibinfo {volume} {106}},\ \bibinfo
  {pages} {093902} (\bibinfo {year} {2011})}\BibitemShut {NoStop}%
\bibitem [{\citenamefont {Liertzer}\ \emph {et~al.}(2012)\citenamefont
  {Liertzer}, \citenamefont {Ge}, \citenamefont {Cerjan}, \citenamefont
  {Stone}, \citenamefont {T\"ureci},\ and\ \citenamefont {Rotter}}]{Lie2012}%
  \BibitemOpen
  \bibfield  {author} {\bibinfo {author} {\bibfnamefont {M.}~\bibnamefont
  {Liertzer}}, \bibinfo {author} {\bibfnamefont {L.}~\bibnamefont {Ge}},
  \bibinfo {author} {\bibfnamefont {A.}~\bibnamefont {Cerjan}}, \bibinfo
  {author} {\bibfnamefont {A.~D.}\ \bibnamefont {Stone}}, \bibinfo {author}
  {\bibfnamefont {H.~E.}\ \bibnamefont {T\"ureci}}, \ and\ \bibinfo {author}
  {\bibfnamefont {S.}~\bibnamefont {Rotter}},\ }\href {\doibase
  10.1103/PhysRevLett.108.173901} {\bibfield  {journal} {\bibinfo  {journal}
  {Phys. Rev. Lett.}\ }\textbf {\bibinfo {volume} {108}},\ \bibinfo {pages}
  {173901} (\bibinfo {year} {2012})}\BibitemShut {NoStop}%
\bibitem [{\citenamefont {Miri}\ \emph
  {et~al.}(2012{\natexlab{a}})\citenamefont {Miri}, \citenamefont
  {Regensburger}, \citenamefont {Peschel},\ and\ \citenamefont
  {Christodoulides}}]{Miri2012_c}%
  \BibitemOpen
  \bibfield  {author} {\bibinfo {author} {\bibfnamefont {M.-A.}\ \bibnamefont
  {Miri}}, \bibinfo {author} {\bibfnamefont {A.}~\bibnamefont {Regensburger}},
  \bibinfo {author} {\bibfnamefont {U.}~\bibnamefont {Peschel}}, \ and\
  \bibinfo {author} {\bibfnamefont {D.~N.}\ \bibnamefont {Christodoulides}},\
  }\href {\doibase 10.1103/PhysRevA.86.023807} {\bibfield  {journal} {\bibinfo
  {journal} {Phys. Rev. A}\ }\textbf {\bibinfo {volume} {86}},\ \bibinfo
  {pages} {023807} (\bibinfo {year} {2012}{\natexlab{a}})}\BibitemShut
  {NoStop}%
\bibitem [{\citenamefont {Miri}\ \emph
  {et~al.}(2012{\natexlab{b}})\citenamefont {Miri}, \citenamefont {Aceves},
  \citenamefont {Kottos}, \citenamefont {Kovanis},\ and\ \citenamefont
  {Christodoulides}}]{Miri2012_b}%
  \BibitemOpen
  \bibfield  {author} {\bibinfo {author} {\bibfnamefont {M.-A.}\ \bibnamefont
  {Miri}}, \bibinfo {author} {\bibfnamefont {A.~B.}\ \bibnamefont {Aceves}},
  \bibinfo {author} {\bibfnamefont {T.}~\bibnamefont {Kottos}}, \bibinfo
  {author} {\bibfnamefont {V.}~\bibnamefont {Kovanis}}, \ and\ \bibinfo
  {author} {\bibfnamefont {D.~N.}\ \bibnamefont {Christodoulides}},\ }\href
  {\doibase 10.1103/PhysRevA.86.033801} {\bibfield  {journal} {\bibinfo
  {journal} {Phys. Rev. A}\ }\textbf {\bibinfo {volume} {86}},\ \bibinfo
  {pages} {033801} (\bibinfo {year} {2012}{\natexlab{b}})}\BibitemShut
  {NoStop}%
\bibitem [{\citenamefont {Joglekar}\ and\ \citenamefont
  {Barnett}(2011)}]{Jog2011}%
  \BibitemOpen
  \bibfield  {author} {\bibinfo {author} {\bibfnamefont {Y.~N.}\ \bibnamefont
  {Joglekar}}\ and\ \bibinfo {author} {\bibfnamefont {J.~L.}\ \bibnamefont
  {Barnett}},\ }\href {\doibase 10.1103/PhysRevA.84.024103} {\bibfield
  {journal} {\bibinfo  {journal} {Phys. Rev. A}\ }\textbf {\bibinfo {volume}
  {84}},\ \bibinfo {pages} {024103} (\bibinfo {year} {2011})}\BibitemShut
  {NoStop}%
\bibitem [{\citenamefont {Vemuri}\ \emph {et~al.}(2011)\citenamefont {Vemuri},
  \citenamefont {Vavilala}, \citenamefont {Bhamidipati},\ and\ \citenamefont
  {Joglekar}}]{Vem2011}%
  \BibitemOpen
  \bibfield  {author} {\bibinfo {author} {\bibfnamefont {H.}~\bibnamefont
  {Vemuri}}, \bibinfo {author} {\bibfnamefont {V.}~\bibnamefont {Vavilala}},
  \bibinfo {author} {\bibfnamefont {T.}~\bibnamefont {Bhamidipati}}, \ and\
  \bibinfo {author} {\bibfnamefont {Y.~N.}\ \bibnamefont {Joglekar}},\ }\href
  {\doibase 10.1103/PhysRevA.84.043826} {\bibfield  {journal} {\bibinfo
  {journal} {Phys. Rev. A}\ }\textbf {\bibinfo {volume} {84}},\ \bibinfo
  {pages} {043826} (\bibinfo {year} {2011})}\BibitemShut {NoStop}%
\bibitem [{\citenamefont {Midya}\ \emph {et~al.}(2010)\citenamefont {Midya},
  \citenamefont {Roy},\ and\ \citenamefont {Roychoudhury}}]{Mid2010}%
  \BibitemOpen
  \bibfield  {author} {\bibinfo {author} {\bibfnamefont {B.}~\bibnamefont
  {Midya}}, \bibinfo {author} {\bibfnamefont {B.}~\bibnamefont {Roy}}, \ and\
  \bibinfo {author} {\bibfnamefont {R.}~\bibnamefont {Roychoudhury}},\ }\href
  {\doibase 10.1016/j.physleta.2010.04.046} {\bibfield  {journal} {\bibinfo
  {journal} {Phys. Lett. A}\ }\textbf {\bibinfo {volume} {374}},\ \bibinfo
  {pages} {2605} (\bibinfo {year} {2010})}\BibitemShut {NoStop}%
\bibitem [{\citenamefont {Graefe}\ and\ \citenamefont {Jones}(2011)}]{Gra2011}%
  \BibitemOpen
  \bibfield  {author} {\bibinfo {author} {\bibfnamefont {E.-M.}\ \bibnamefont
  {Graefe}}\ and\ \bibinfo {author} {\bibfnamefont {H.~F.}\ \bibnamefont
  {Jones}},\ }\href {\doibase 10.1103/PhysRevA.84.013818} {\bibfield  {journal}
  {\bibinfo  {journal} {Phys. Rev. A}\ }\textbf {\bibinfo {volume} {84}},\
  \bibinfo {pages} {013818} (\bibinfo {year} {2011})}\BibitemShut {NoStop}%
\bibitem [{\citenamefont {Szameit}\ \emph {et~al.}(2011)\citenamefont
  {Szameit}, \citenamefont {Rechtsman}, \citenamefont {Bahat-Treidel},\ and\
  \citenamefont {Segev}}]{Sza2011}%
  \BibitemOpen
  \bibfield  {author} {\bibinfo {author} {\bibfnamefont {A.}~\bibnamefont
  {Szameit}}, \bibinfo {author} {\bibfnamefont {M.~C.}\ \bibnamefont
  {Rechtsman}}, \bibinfo {author} {\bibfnamefont {O.}~\bibnamefont
  {Bahat-Treidel}}, \ and\ \bibinfo {author} {\bibfnamefont {M.}~\bibnamefont
  {Segev}},\ }\href {\doibase 10.1103/PhysRevA.84.021806} {\bibfield  {journal}
  {\bibinfo  {journal} {Phys. Rev. A}\ }\textbf {\bibinfo {volume} {84}},\
  \bibinfo {pages} {021806} (\bibinfo {year} {2011})}\BibitemShut {NoStop}%
\bibitem [{\citenamefont {El-Ganainy}\ \emph {et~al.}(2012)\citenamefont
  {El-Ganainy}, \citenamefont {Makris},\ and\ \citenamefont
  {Christodoulides}}]{Elg2012}%
  \BibitemOpen
  \bibfield  {author} {\bibinfo {author} {\bibfnamefont {R.}~\bibnamefont
  {El-Ganainy}}, \bibinfo {author} {\bibfnamefont {K.~G.}\ \bibnamefont
  {Makris}}, \ and\ \bibinfo {author} {\bibfnamefont {D.~N.}\ \bibnamefont
  {Christodoulides}},\ }\href {\doibase 10.1103/PhysRevA.86.033813} {\bibfield
  {journal} {\bibinfo  {journal} {Phys. Rev. A}\ }\textbf {\bibinfo {volume}
  {86}},\ \bibinfo {pages} {033813} (\bibinfo {year} {2012})}\BibitemShut
  {NoStop}%
\bibitem [{\citenamefont {Zezyulin}\ and\ \citenamefont
  {Konotop}(2012)}]{Zez2012}%
  \BibitemOpen
  \bibfield  {author} {\bibinfo {author} {\bibfnamefont {D.~A.}\ \bibnamefont
  {Zezyulin}}\ and\ \bibinfo {author} {\bibfnamefont {V.~V.}\ \bibnamefont
  {Konotop}},\ }\href {\doibase 10.1103/PhysRevLett.108.213906} {\bibfield
  {journal} {\bibinfo  {journal} {Phys. Rev. Lett.}\ }\textbf {\bibinfo
  {volume} {108}},\ \bibinfo {pages} {213906} (\bibinfo {year}
  {2012})}\BibitemShut {NoStop}%
\bibitem [{\citenamefont {Miri}\ \emph
  {et~al.}(2012{\natexlab{c}})\citenamefont {Miri}, \citenamefont {LiKamWa},\
  and\ \citenamefont {Christodoulides}}]{Miri2012_a}%
  \BibitemOpen
  \bibfield  {author} {\bibinfo {author} {\bibfnamefont {M.-A.}\ \bibnamefont
  {Miri}}, \bibinfo {author} {\bibfnamefont {P.}~\bibnamefont {LiKamWa}}, \
  and\ \bibinfo {author} {\bibfnamefont {D.~N.}\ \bibnamefont
  {Christodoulides}},\ }\href {\doibase 10.1364/OL.37.000764} {\bibfield
  {journal} {\bibinfo  {journal} {Opt. Lett.}\ }\textbf {\bibinfo {volume}
  {37}},\ \bibinfo {pages} {764} (\bibinfo {year}
  {2012}{\natexlab{c}})}\BibitemShut {NoStop}%
\bibitem [{\citenamefont {Kocinac}\ \emph {et~al.}(2005)\citenamefont
  {Kocinac}, \citenamefont {Milanovic}, \citenamefont {Ikonic}, \citenamefont
  {Indjin}, \citenamefont {Radovanovic},\ and\ \citenamefont
  {Harrison}}]{Koc2005}%
  \BibitemOpen
  \bibfield  {author} {\bibinfo {author} {\bibfnamefont {S.}~\bibnamefont
  {Kocinac}}, \bibinfo {author} {\bibfnamefont {V.}~\bibnamefont {Milanovic}},
  \bibinfo {author} {\bibfnamefont {Z.}~\bibnamefont {Ikonic}}, \bibinfo
  {author} {\bibfnamefont {D.}~\bibnamefont {Indjin}}, \bibinfo {author}
  {\bibfnamefont {J.}~\bibnamefont {Radovanovic}}, \ and\ \bibinfo {author}
  {\bibfnamefont {P.}~\bibnamefont {Harrison}},\ }\href {\doibase
  10.1002/pssc.200461823} {\bibfield  {journal} {\bibinfo  {journal} {phys.
  stat. sol. (c)}\ }\textbf {\bibinfo {volume} {2}},\ \bibinfo {pages} {3552}
  (\bibinfo {year} {2005})}\BibitemShut {NoStop}%
\bibitem [{\citenamefont {Arfken}\ and\ \citenamefont
  {Hans}(2001)}]{Arfken2001}%
  \BibitemOpen
  \bibfield  {author} {\bibinfo {author} {\bibfnamefont {G.~B.}\ \bibnamefont
  {Arfken}}\ and\ \bibinfo {author} {\bibfnamefont {J.~W.}\ \bibnamefont
  {Hans}},\ }\href@noop {} {\emph {\bibinfo {title} {Mathematical Methods For
  Physicists}}},\ \bibinfo {edition} {5th}\ ed.\ (\bibinfo  {publisher}
  {Academic},\ \bibinfo {address} {San Diego},\ \bibinfo {year}
  {2001})\BibitemShut {NoStop}%
\end{thebibliography}%

\end{document}